\documentclass[aps,preprint,showpacs]{revtex4-1}
\usepackage{dcolumn}
\usepackage{bm}
\usepackage{amssymb}
\usepackage{amsfonts}
\usepackage{amsmath}
\usepackage{graphicx}

\begin{document}

\title{Estimating the number of tissue resident macrophages}
\author{Augusto Gonzalez}
\affiliation{Instituto de Cibernetica, Matematica y Fisica, La Habana, Cuba}

\begin{abstract}
I provide a simple estimation for the number of macrophages in a tissue, arising from the hypothesis
that they should keep infections below a certain threshold, above which neutrophils are 
recruited from blood circulation. The estimation reads $N_m=a N_{cel}^{\alpha}/N_{max}$, 
where $a$ is a numerical coefficient, the exponent $\alpha\approx 2/3$, and $N_{max}$ is the
maximal number of pathogens a macrophage may engulf in the time interval, $t_r$, between
pathogen replications.
\end{abstract}

\maketitle

Tissue resident macrophages are a subject under intense research by the scientific 
community \cite{trm}. The estimation of their numbers in different tissues is a key
problem in the understanding of how the immune system works just after pathogens 
arrive to a given tissue. A serious effort in determining the average number of all
cells of the immune system has already been initiated \cite{mapT}. In my opinion, 
the only drawback of such an effort is that there is no idea of what numbers one
should expect.

In the present paper, I use very simple reasonings in order to obtain an estimation 
for the number of resident macrophages in a tissue with $N_{cel}$ cells.

I start by considering the free evolution of pathogens (bacteria, for example) in a 
tissue. In the initial instants, the number of pathogens follows an exponential law:

\begin{equation}
N_p=N_p(0)~ 2^{t/t_r}=N_p(0) \exp (t \ln (2)/t_r),
\label{eq1}
\end{equation}

\noindent
where $N_p(0)$ is the number at $t=0$, and $t_r$ is the replication time. The latter is
of the order of one hour for bacteria \cite{bactT}.

\begin{figure}[ht]
\begin{center}
\includegraphics[width=0.9\linewidth,angle=0]{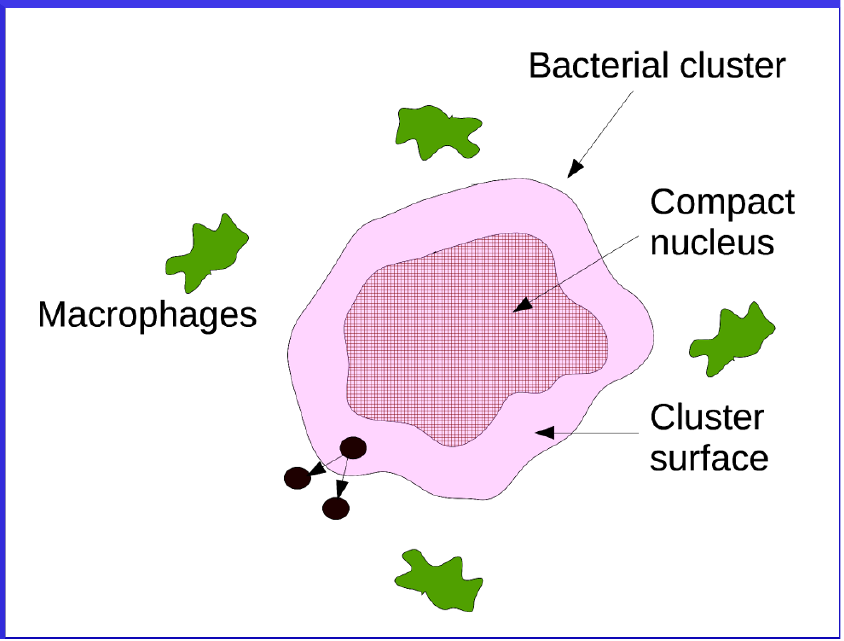}
\caption{Evolution of a bacterial cluster at later stages, when only bacteria in the 
surface have enough space to undergo mitosis. The cluster is surrounded by macrophages.}
\label{fig1}
\end{center}
\end{figure}

The exponential growth is, however, constrained by geometry. At later times, the pathogen 
cluster becomes compact, and only bacteria at the surface may divide, as shown in Fig. 
\ref{fig1}. The number of new pathogens in a time interval $t_r$ is, thus, proportional to
the cluster surface:

\begin{equation}
\Delta N_p\approx a' N_p^{2/3}.
\label{eq2}
\end{equation}

\noindent
Indeed, the cluster radius is $R\sim N_p^{1/3}$, and its surface $R^2 \sim N_p^{2/3}$.

With regard to macrophages, I assume that they are uniformly distributed in the 
tissue, and exhibit high motility. As $t_r$ is large enough, if the bacterial cluster
grows above certain limits, in a time interval of about 3 or 4 $t_r$  most of macrophages can
be mobilized to the site of infection. Immunity in a tissue requires that the number of 
new pathogens, Eq. (\ref{eq2}), should be lower than those destroyed by macrophages.
The latter is $N'_m N_{max}$, where $N'_m$ is the number of macrophages that have
already arrived to the site of infection, and $N_{max}$ is the maximal number of
pathogens that a macrophage may engulf in the time interval $t_r$. Thus:

\begin{equation}
a' N_p^{2/3} < N'_m N_{max} < N_m N_{max}, 
\label{eq3}
\end{equation}

\noindent
where $N_m$ is the number of macrophages in the tissue.

Our final expression for $N_m$ comes from the idea of a threshold for $N_p$, above
which neutrophils are called to help fighting the infection. The number of injured 
cells is naturally related to $N_p$. And these injured cells should not overcome
a fraction of the number of cells in the tissue (let's say, one hundredth of them,
for example). Then, we can write $N_{cel}$ instead of $N_p$ in Eq. (\ref{eq3}), and
introduce a new coefficient $a$, instead of $a'$, arriving to:

\begin{equation}
N_m > \frac{a N_{cel}^{\alpha}}{N_{max}}. 
\label{eq4}
\end{equation}

This is the main result of the paper. The exponent following from Eq. (\ref{eq3}) is
$\alpha=2/3$, but I have written it more generally in order to consider the effect of
different tissue effective dimensionality. With regard to $N_{max}$, I guess that it
takes similar values for all of the tissues. In around one hour time, a single macrophage 
in the hyper-activated state may destroy 50 or even a larger number of bacteria \cite{nmax},
for example.

\begin{figure}[ht]
\begin{center}
\includegraphics[width=0.9\linewidth,angle=0]{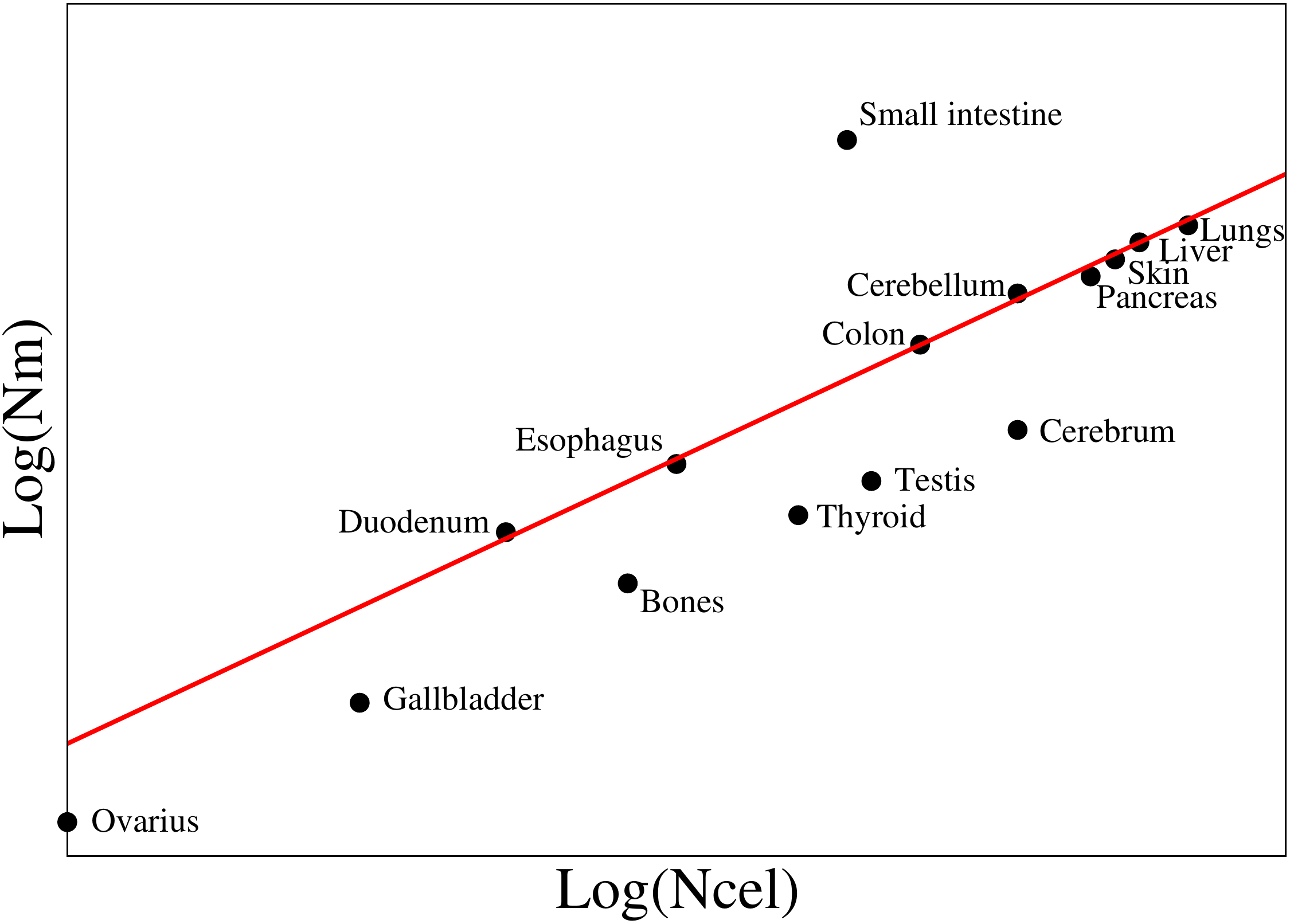}
\caption{Schematics of $\log (N_m)$ vs $\log (N_{cel})$ in tissues.}
\label{fig2}
\end{center}
\end{figure}

A schematics of what I expect is represented in Fig. \ref{fig2} for the tissues analized in
Ref \cite{picontrol}. For simplicity, they are labelled by the organ in which they reside.
I rewrite Eq. (\ref{eq4}) as $\log (N_m)>\alpha\log (N_{cel})+\log (a/N_{max})$. In a log-log 
plot of $N_m$ vs $N_{cel}$, I expect a set of tissues to be grouped along a line with slope 
$\alpha\approx 2/3$ (a red line in Fig. \ref{fig2}).
These are the ``normal'' tissues \cite{picontrol}. The coefficient $a$ is very similar for all
of them. This coefficient is related to the threshold fraction of tissue that is allowed to be 
injured by the pathogens.

Precisely due to lower values of $a$, tissues with ``reduced'' (or privileged) immune protection
are located below the line of normal tissues. This reduction in the number of macrophages is 
compensated by physical barriers (cerebrum, testis) or by physiological conditions (high ph of
bile in the gallbladder), for example.

On the other hand, I expect at least one tissue well above the line: the small intestine. This 
time not $a$, but what is higher than normal is the average number of pathogens arriving to
the distal end of the small bowel \cite{picontrol}. Eq. (\ref{eq4}) gives a lower bound for $N_m$.
In the small intestine, $N_m$ should be much larger than its lower bound in order to protect
the tissue against pathogen overload.

In conclusion, I suggest that the number of macrophages resident in a tissue is proportional to
$N_{cel}^{\alpha}$, where $\alpha\approx 2/3$. A group of tissues should follow this law (the normal
tissues). In addition, there should be a second group located below the line of normal tissues, and
at least one tissue, the small intestine, above that line. The hypothesis can be tested in the
near future.

{\bf Acknowledgments.}
Support from the National Program of Basic Sciences in Cuba, and from the Office of External 
Activities of the International Center for Theoretical Physics (ICTP) is acknowledged.

\end{document}